# Physical-mechanical characterization of hydraulic and non-hydraulic lime based mortars for a French porous limestone

M. AL-MUKHTAR[1]* & K. BECK[1,2]

[1]*Centre de Recherche sur la Matière Divisée, Ecole Polytechnique de l'Université d'Orléans, CNRS - CRMD, 1B rue de la Férollerie, 45071 Orléans Cedex 2, France.*
[2]*Institut des Sciences de la Terre d'Orléans, CNRS-ISTO, 1A rue de la Férollerie, 45071 Orléans Cedex 2, France.*

* muzahim@cnrs-orleans.fr

ABSTRACT: The focus of the study presented in this paper is to provide reliable criteria that can be used to estimate the degree of compatibility between the French limestone tuffeau and mortar. It is suggested through this study to use the same parent material (i.e., tuffeau) as mortar. The mortar used in this study is composed of non-hydraulic (hydrated) lime or hydraulic lime and aggregates obtained from fragments and powder of the tuffeau stone. Water transfer properties and mechanical behaviour of the mortars are evaluated and compared with the original stone Tuffeau. Based on these studies, some key guidelines are provided such that a mortar that is compatible with properties of Tuffeau and can be prepared and used as construction material of monuments and maintenance purposes.

## 1 INTRODUCTION

Sedimentary rocks such as limestones have been commonly used in the construction of several historically important monuments and structures for many centuries in various parts of the world. These structures are standing examples of our cultural heritage and civilization. For this reason, such structures should be preserved and well maintained for our future generations. However, due to the action of water and the influence of environment and pollution, these structures gradually deteriorate over a period of time. Several millions of dollars are annually spent for the maintenance and repair works of these valuable structures in various parts of the world. There are various construction methods that have been developed based on experience to repair these structures. Investigations in this direction has shown suitable mortars can be suggested for repair works by better understanding the mechanical, physical and chemical properties of the mortars in addition to a rigorous evaluation process of their performance under varying environmental conditions. Analysis of numerous cases of degradation of the architectural monuments in Loire Valley (France) show that the damages often find their origin to the type of mortar that is used in the construction. From the second half of the 20[th] century, restoration of several historical monuments was undertaken mainly using cement mortars as joints between stones. Several damages were attributed due to the use of cement mortar particularly when used with limestones (Rautureau 2001). Studies have shown chemical characteristics of cement (essentially its alkaline salts content) are incompatible with limestone. Furthermore, cement mortars are hard, stiff and have a high adhesion force which partly contributes towards the increase in the brittleness property of stones. In addition, an impermeable mortar favors the accumulation of water in the stone and constitutes a zone that increases dissolution and recrystallization (Fig. 1).

Lime based mortars are more commonly used since ancient times in the construction (Egyptians, Chinese, Phoeniciens) (Furlan & Bisseger 1975; Biscontin et al. 2002, Fassina et al. 2002). Conventionally, two types of lime are used as mortars (i) non-hydraulic lime (also called hydrated lime) made from relatively pure limestones which mainly harden due to carbonation and (ii) natural hydraulic lime made from limestone with reactive silica and aluminium impurities which harden and set in the presence of water.

This paper presents the behaviour of lime and examines their compatibility with the tuffeau, which is a typical porous limestone which is available plentiful in the Loire Valley (France). The prepared mortar is composed with non-hydraulic or hydraulic limes and aggregates (powder) obtained from the stone tuffeau. Mortars prepared using the same material (i.e., tuffeau) will have similar physico-chemical and hydro-mechanical properties and compatible with those of the tuffeau stone. Mechanical behavioural properties (strength properties in compression and tension) and hydraulic property (mainly capillary imbition) were evaluated for mortar samples composed with different proportions of lime.

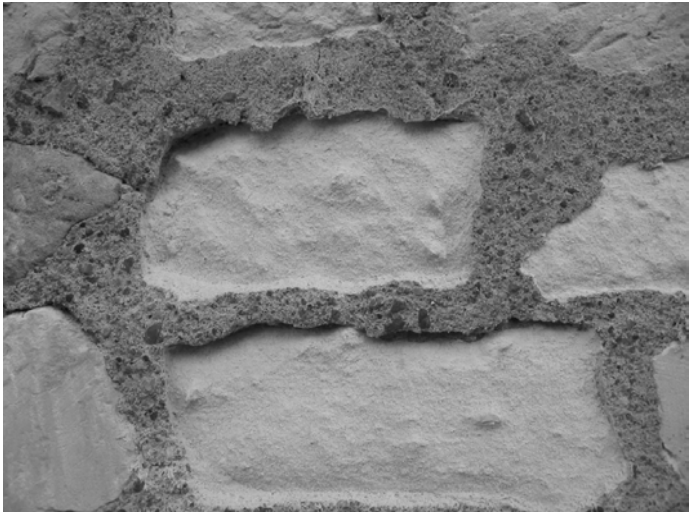

Figure 1: Deterioration of a tuffeau stone associated with the use of cement joint mortar

## 2. MATERIALS AND METHODS

### 2.1 Mineralogical and textural characterisation of the building stone

The Tuffeau stone used in the study is a porous sedimentary white limestone (Table 1). The principal minerals present in it are calcite ($CaCO_3$) and silica ($SiO_2$) in the form of opal cristobalite-tridymite and quartz. The other minerals present in the stone are micas, clays and detritic minerals such as $TiO_2$. The texture of stones forms complex porous network (Beck et al., 2003). The arrangement of larger and smaller grains within the stones contributes to macropores and micropores within the stone structure.

Table 1: Main characteristics of the Tuffeau (Beck et al., 2003).

| Parameter | Tuffeau |
| --- | --- |
| Mineralogical composition | Calcite ≅ 50%, Opal CT ≅ 30%, Quartz ≅ 10%, Clay and Mica ≅10% |
| Skeletal density (g/cm³) | 2.55 |
| Bulk dry density (g/cm³) | 1.31 |
| Porosity | ≅ 48% |

### 2.2 Mortar components characterisation

The mortar is composed of Tuffeau limestone powder and lime. The powder of Tuffeau is obtained by crushing of the stone in a grinder-mixture followed by sifting to achieve a fine powder by passing through a sieve of 315 μm diameter. The hydraulic and non-hydraulic lime powders that are used are light and have low densities. Figure 2 shows the particle size distributions obtained from laser granulometry of Tuffeau powder of the two limes used in this study. From this figure it can be seen that Tuffeau particles are fine and well graded. Hydraulic lime particles are very close to that of Tuffeau. Non-hydraulic lime particles are less than 50 μm in size and are smaller than Tuffeau particles. However, the dimensions of all particles that are used in the preparation of the mortar are relatively small with sizes ranging between 1 to 100 μm.

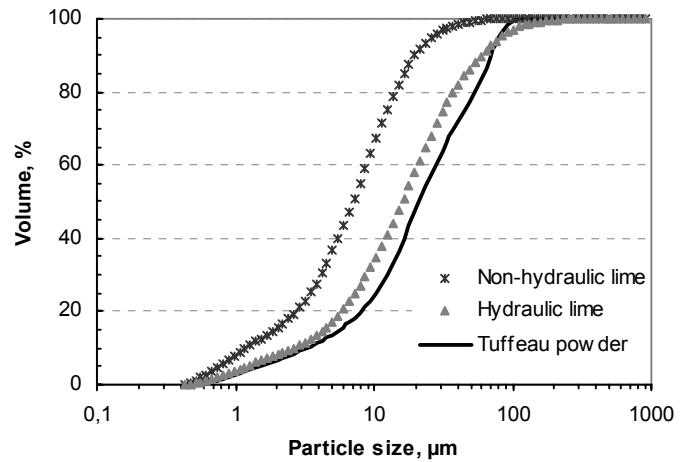

Figure 2: Particle size distribution of powders from laser granulometry

### 2.3 Mortar preparation

The test specimens of mortar consisted of only two components: powder of Tuffeau and lime (hydraulic and non-hydraulic lime). The effect of varying the concentration of lime (i.e. 5 – 50%) in the mortar paste was studied. Water content of the mortar pastes was set at 50% so that the mortar paste would have a good workable consistency no matter what the concentration of lime. Two different types of samples were prepared. In the first set, prepared samples are 80 mm in height and 40 mm in diameter. They were compacted in a cylindrical cast of 40 mm in diameter under a specific load in order to achieve desired dry bulk density at 1.2 g/cm³ (close to that of the stone) and allowed to set for two days. The specimens were then remoulded and placed in a hermitic chamber at a relative humidity of 100% and constant temperature of 20°C for 25 more days in order to avoid desaturation phenomena. And finally, they were dried in an oven at 105°C for a period of

24 hours. For this reason, the specimens were allowed to harden for 28 days before being tested.

In the second series (used only for the adhesion tests) a mortar of 10 mm of thickness is placed on cylindrical stone sample (diameter: 40 mm and height: 35 mm) without compaction. During setting, the mixture of stone/mortar is placed in a bag and sealed for a period of 7 days in order to prevent evaporation. Then, samples are unpacked and kept in a desiccator (at 76% relative humidity and 20°C temperature) for a period of 21 days before testing.

*2.4 Mechanical tests*

A number of mechanical tests (Beck & Al-Mukhtar, 2006) were carried out (Fig. 3) on cylindrical samples of mortar which include: compressive strength (Standard AFNOR P94-420) and tensile strength called also as cylinder splitting test (Standard AFNOR P94-422). In addition adhesion test (Standard AFNOR NF EN 1015-12) was conducted on the mortar paste placed on cylindrical stone sample without compaction. An increasing force is applied at a loading rate of 0.05 mm/min for compressive and tensile tests. The samples were loaded at a rate of 0.5 MPa/min using an Instron 4485 press to determine the rupture in the tested samples (adhesive test).

The volumetric deformation behaviour of the samples was also determined. The reduction in volume of initial saturated samples when subjected to a relative humidity of 32% (humidity induced by using a saturated salt solution in a desiccator's) was determined. These tests are referred as "preliminary shrinkage tests" in a later section in the paper. This terminology was used as authors feel that more testing is required in this direction.

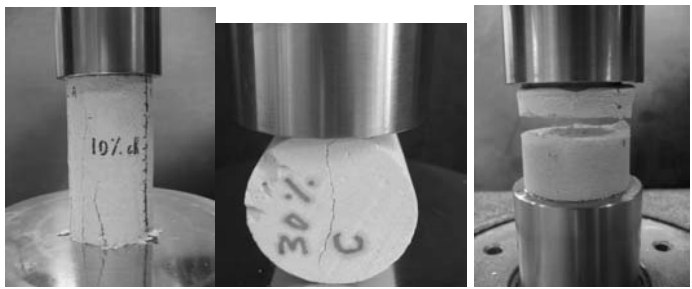

Figure 3: Mortar samples during compressive, tensile and adhesive tests respectively

*2.5 Water absorption-imbibition test*

In this test (standard AFNOR B10-613), previously dried cylindrical samples are placed in a hermetic tank at the bottom of which distilled water level is maintained constant during entire period of the test (Fig. 4). The mass of wet samples and the height of the capillary front are measured at increasing intervals of times. The imbibition coefficients *A* and *B* correspond to the slopes of the curves of mass uptake and rise of the capillary front according to the square root of elapsed time.

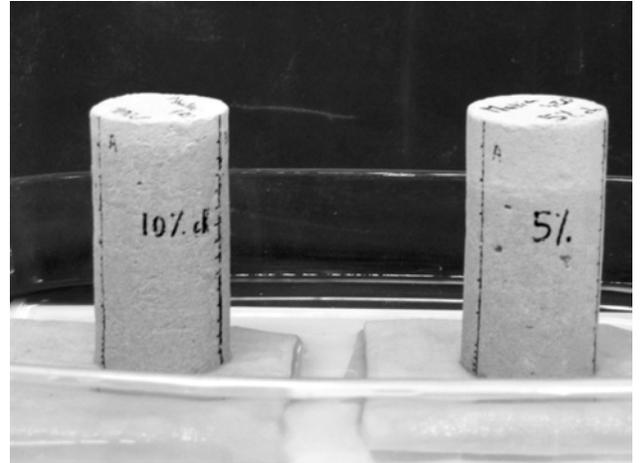

Figure 4: Mortar samples during imbibition tests

## 4. RESULTS

*4.1 Compressibility of the mortar*

One dimensional compression tests (oedometer tests – Fig. 5) were performed on mortar pastes (lateral deformation are prevented during the test) to determine its compressibility behaviour under drained conditions. The mortar paste was prepared using the different concentrations of lime at a solid/liquid ratio of 1:1. The duration of this test is about 2-3 hours which is much lower than the initial setting time of the tested paste (> 10 hours). The sample pastes are placed in 65 mm diameter oedometer specimen rings. The axial stress is applied and is maintained until axial strain stabilizes. At the end of the loading, the height of sample is measured and dry bulk density and porosity are determined from the initial properties of the test specimen.

Results are presented in terms of void ratio (e) versus axial applied stresses (Fig. 5). The results demonstrate that the void ratio reduces as the axial stresses are increased. During unloading, the void ratio increases but does not go back to its initial value. The hysteretic behaviour demonstrates that the material is not elastic but elasto-plastic and each action of loading affects the structure of the mortar permanently. The compression index, $C_c$ is defined as the changes in void ratio ($\Delta e$) per logarithmic cycle of axial vertical stress ($\sigma_v$):

$$Cc = \frac{-\Delta e}{\Delta(\log \sigma_v)}$$

Results indicate (table 2) that the hydraulic and non-hydraulic lime based mortar compression index values are similar for all tested proportions of lime in the mortar. In other words, the effect of different percentages of lime is negligible on the compressibility behaviour of the mortar. Such behaviour can be explained by the difference in the particle size distribution of the two limes and Tuffeau grain powders.

In order to ensure a similar hydraulic behaviour between mortar and tuffeau stone, samples are compacted to dry bulk density close to that of tuffeau (i.e. 1.2 g/cm$^3$). Table 3 presents the porosity of past mortar samples compacted at this density. Porosity values are determined from void ratio values as:

$$\text{Porosity} = Void\ ratio / (1 + Void\ ratio)$$

Values of porosity obtained are close to that of tuffeau which is 0.48.

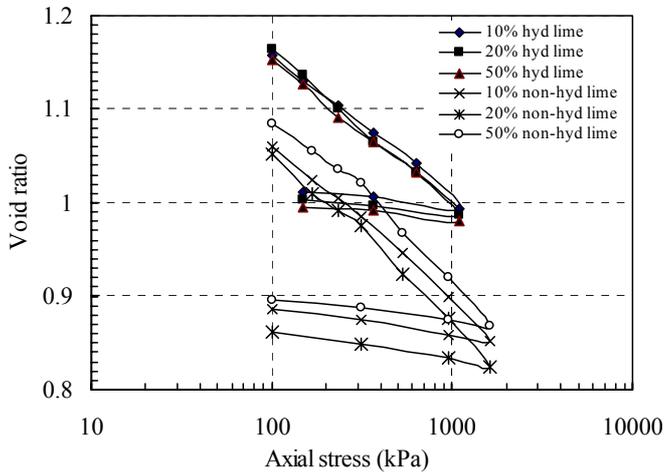

Figure 5: One dimensional compression tests on the past mortar samples

Table 2: Compression index of tested mortar

| Mortar joint with | Compression index $C_c$ | |
| --- | --- | --- |
| | Non-hydraulic lime | Hydraulic lime |
| 10% lime | 0.16 | 0.16 |
| 20% lime | 0.17 | 0.175 |
| 50% lime | 0.172 | 0.17 |

Table 3: Porosity of mortar pastes for a fixed dry bulk density

| Porosity in the mortar compacted to $\gamma_d$ = 1.20 g/cm$^3$ | 10% lime | 20% lime | 30% lime | 50% lime |
| --- | --- | --- | --- | --- |
| Non-hydraulic mortar | 0.51 | 0.51 | 0.50 | 0.49 |
| Hydraulic mortar | 0.52 | 0.52 | 0.53 | 0.53 |

*4.2 Mechanical properties*

The mechanical behaviour of mortar specimens was examined by comparing tests results of compressive strength, tensile strength and adhesive strength. For tuffeau stone at the dry state, the compressive strength is 12 MPa and the tensile strength is 1.5 MPa. Compressive strength results as presented in Figure 6 increases with the lime concentration. Meanwhile, mortar samples prepared with hydraulic lime have higher mechanical strength than mortars prepared with non-hydraulic lime. Tensile strength results as presented in Figure 7. Mortars are weaker in tension than in compression: tensile strength is less than one tenth of compressive strength. However, the rate of increase of strength in compressive strength is much higher in comparison to the tensile strength. The reaction of hydraulic lime with tuffeau powder is quicker than that non-hydraulic lime. The non-hydraulic lime based mortar set is realized by pozzolanic reation with stone aggregates (clayey minerals) and also by carbonisation reaction with $CO_2$ from air. This reaction is very slow and may need several months to achieve an equivalent strength to that of hydraulic lime.

For the adhesive strength, results obtained after 28 days of curing (Table 4) shows that adhesive strength increases as lime percentage increases. These values can be considered as acceptable for lime based mortar (Bromblet, 2000).

Figure 8 shows preliminary test results of shrinkage tests. The test results suggest volumetric deformation variations stabilise practically within 2 weeks for all tested mortars. The volumetric variations determined were less than 10 % in samples prepared with hydraulic and non-hydraulic lime. Changes in the behaviour due to the continuous reaction between lime and aggregates (tuffeau powder) occurred without changes in water content after 2 weeks of curing.

These tests have been carried out on few samples and so more samples must be tested to identify the exact difference in the behaviour between the two tested limes. Furthermore, based on the mechanical behaviour results, lime mortars must have a lime concentration of at least 20% to be effective, but it is important to note that mechanical parameters will be improved with time (Lanas and Alvarez, 2003).

Table 4: Adhesive strength for the tested mortars

| Mortar joint with | Adhesive strength (kPa) | |
| --- | --- | --- |
| | Non-hydraulic lime | Hydraulic lime |
| 5% lime | 0.0 | 0.0 |
| 10% lime | 47 | 72 |
| 20% lime | 119 | 94 |
| 30% lime | 205 | 124 |
| 50% lime | Not measured | 240 |

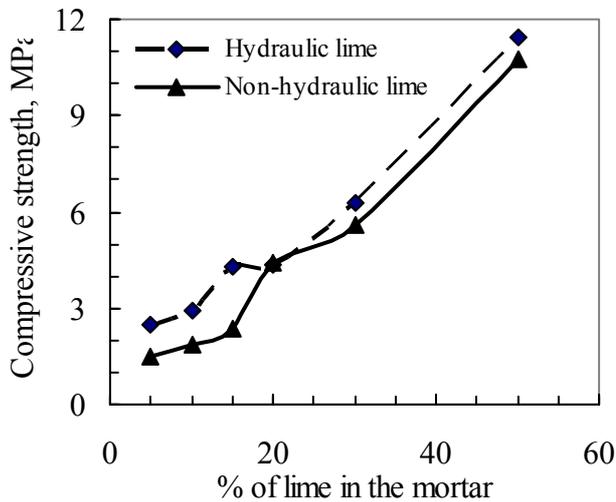

Figure 6: Compressive strength of the mortar after 28 days of curing

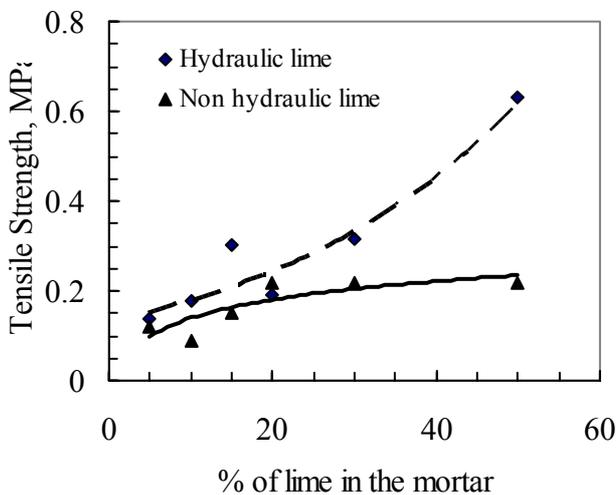

Figure 7: Tensile strength of the mortar after 28 days of curing

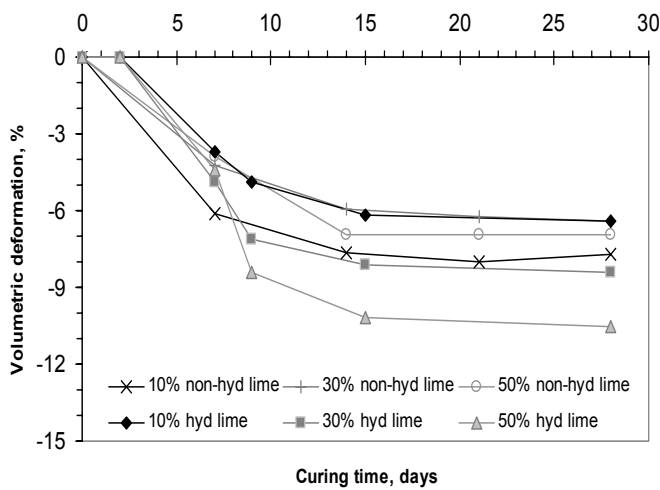

Figure 8: Volumetric deformation of the mortar during 28 days of curing

*4.3 Capillary water absorption property*

Water transfer tests by capillary absorption measurements were performed on each of the test samples before mechanical strength tests were determined. Figure 9 shows imbition coefficient of mass uptake (A) and capillary ascent (B) for different tested mortars. For tuffeau stone imbition coefficient are: A=0.36 (g/cm$^2$/min$^{1/2}$) and B=0.96 (cm/min$^{1/2}$). The main observations are:
- imbition coefficients for mortars prepared with low lime concentrations are close to that of Tuffeau,
- little difference between hydraulic and non-hydraulic lime mortars mainly for lime ≥ 15%.

Irrespective of the lime contents, the total porosity of different prepared mortars was found to be higher than the total porosity of tuffeau stone. The difference in the imbition results can mainly be attributed to the pore size distribution. This behaviour indicates clearly that capillary water absorption properties in a porous material are directly related to the pore network characteristics: pore sizes and shapes. In addition, the following observations can be summarized based on comparing imbition test results to mechanical test results:
- to approach the mechanical strength of tuffeau, mortars must have high lime content ≥ 20%
- to approach the hydraulic behaviour (imbition, characteristics of Tuffeau), mortars must have low lime content ≤ 20%.

The contradictory behaviour observed in this study can be explained by the fact that measurement are carried out only after 28 days. It is well known that the reaction between lime and clayey minerals occurs over a long period of time. For this reason, it can be stated with certain degree of confidence that the mechanical strength will be improved with time. The research studies summarized in this paper suggest that lime percentage about 20% can be sufficient and compatible to tuffeau stone.

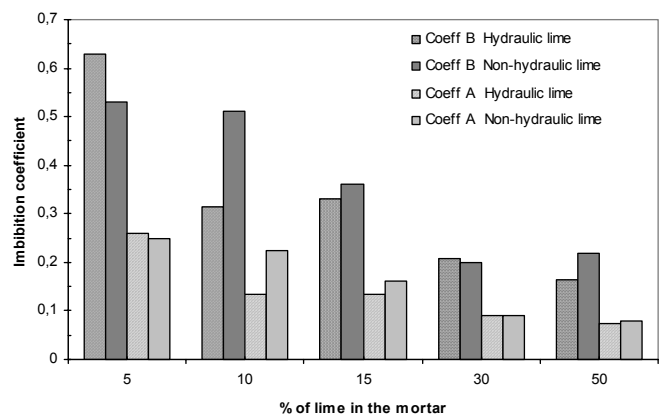

Figure 9: Imbition coefficients for the different mortar

## 5. CONCLUSION

The durability characteristics both in construction or restoration of monuments are dependent on the use of proper mortar. Results obtained and several studies reported in the literature (Al-Saad & Abdel-Halim 2001; Degryse et al. 2002; Binda et al. 2003; Henriques 2005) suggest that the following parameters must be considered in a mortar-stone compatibility study: chemical, mechanical and hydraulic properties. It is more practical to propose a lime based mortar with aggregates originated from the limestone in order to have an excellent compatibility with the chemical properties. The study suggests that mortar seems to have similar mechanical strength properties and is found to be compatible to that of tuffeau stone mainly if lime-aggregate reaction is taken into consideration over a longer period of time. The hydraulic properties need to be further investigated more mainly with respect to permeability and diffusion variations. However, imbibition results indicate mortars using low percentages of lime ($\leq$ 20%) are compatible with tuffeau.

Moreover, this study demonstrates clearly that there is no difference in the behaviour of the mortar composed of hydraulic or non-hydraulic lime. However, these tests have been carried out on few samples and so more samples must be tested to for better understanding of differences in the behaviour between the two tested limes. Microscopic observation and mercury porosimetry investigation studies are necessary for different samples of mortar in order to follow and to compare the microstructure and the pore size distribution changes with lime type and proportion.

Furthermore, aging tests including thermo-hydro-mechanical cycles under real environmental conditions are necessary to evaluate the performance (deterioration and durability) of the proposed mortar in order to confirm the compatibility of the conceived mortar (based on hydraulic or non-hydraulic) with the tuffeau.

## ACKOWLEDGEMENTS

The authors would like to express their thanks to:
- the quarry of tuffeau LUCET
- Dr. Sai Vanapalli, (University of Ottawa, Canada) for his helpful comments on the paper and
- Mr Philippe Badets assistance in the development experimental apparatus for conducting this testing program is appreciated.